\documentstyle[epsf]{l-aa}

\begin{document}
\topmargin=1.0cm
\thesaurus{11.05.2; 11.06.2; 11.09.2; 11.16.1}
\title{Morphology Transformation in Pairs of 
Galaxies-- The Local Sample}

\author{Selma Junqueira\inst{1}, Du\'\i lia F. de 
Mello\inst{1}\thanks{{\em Present address:} STScI, Baltimore 
MD21218} \and Leopoldo Infante\inst{2}}
\institute{Observat\'orio Nacional - C. Postal 23002, 
Rio de Janeiro, 20921-400 Brazil 
\and
       Departamento de Astronom\'\i a y Astrof\'\i sica,
       P. Universidad Cat\'olica de Chile, Casilla 104, 
       Santiago 22, Chile}
\date{Received ; Accepted }
\offprints{selma@on.br}
\maketitle
\markboth{Pairs of Galaxies}{Junqueira, 
de Mello, Infante}

\begin{abstract}

We present photometric analysis of a local sample 
of 14 isolated pairs of galaxies. The photometric 
properties  analyzed in the local pairs are: colors, 
morphology, tidal effects  and activity.  We  verify  
that  close pairs  have  an   excess  of early--type 
galaxies and many  elliptical galaxies in this pairs 
are, in fact, lenticular galaxies.  Many late--pairs 
in our sample show strong   tidal  damage  and  blue 
star formation regions. We conclude that pairs of 
different morphologies may have passed through different 
evolution processes which violently transformed their 
morphology. Pairs with at least one early--type 
component may be descendents of groups of galaxies. 
However, late--type pairs are probably long--lived 
showing clearly signs of interaction. Some of them 
could be seen as an early stage of mergers. These 
photometric database will be used for future comparison 
with more distant pairs in order to study galaxy 
evolution.

\end{abstract}

\section{Introduction}

Since Hubble (1926, 1936) categorized galaxies according 
to their shapes, galaxy morphology has been used as a 
fundamental parameter in studies of galaxy formation and 
evolutionary processes. However, many galaxies have unusual 
morphologies and do not belong to any Hubble type. Most of 
these peculiar galaxies are interacting systems and had 
their forms distorted by tidal forces. Therefore, 
interactions between galaxies are a major evolution 
mechanism which transform galaxy morphology during their 
evolution. The so-called Toomre's sequence (Toomre 1977) 
is a perfect example of what happens to galaxy morphology 
during different stages of interaction. Numerical 
simulations of the encounter between galaxies has 
established that gravitational interaction can not only 
originate tidal damage to spiral structures but also merge 
two disk galaxies in one final product which resembles an
elliptical galaxy (Toomre \& Toomre 1972, see Barnes 
\& Hernquist 1992 for a review). In order to establish 
whether an encounter will originate a merger or not
it is necessary to determine the physical properties and 
dynamics of the interaction. In principle, all systems 
made of galaxies with comparable mass and which are
in elliptical bound orbits end as a merger but not 
necessarily within the Hubble time (Binney \& Tremaine 
1987, Junqueira \& de Freitas Pacheco 1994). 

If mergers are frequent in galaxy evolution we should be 
able to find mergers in progress such as NGC7252 and/or 
traces of the effect produced by this phenomenon.
For instance, the most commonly associated features invoked 
by merger theory are bars, isophotes twisting, color 
gradient, shells, inner structures, dust, starbursts, 
AGN's, etc (Schweizer 1983, Noguchi 1988, Bender et al. 
1989, Forbes \& Thomson 1992). However, to what degree is 
the galaxy properties influenced by interactions with its 
neighbors is still an open issue. For instance, interaction 
involving at least one gas-rich galaxy may result in 
activity which may range from global starbusts to nuclear 
ones and to AGN's, however it is unclear whether there is 
a transition between these three features (Fritze-v. 
Alvensleben 1996, Keel 1996).

One of the most striking evidence for galaxy evolution is 
the faint blue galaxies excess at $z \ge 0.3$, i.e. the 
Buchter \& Oemler effect (Butcher \& Oemler 1978, 1984). In 
order to explain this effect, galaxy evolution models need 
a mechanism for creating disturbed galaxies with enhanced 
star formation and explain why they occur more frequently at 
this redshift (Moore et al. 1996). Moreover, they have to 
address the question of morphology evolution and an 
identification of the remnants of the distorted blue 
galaxies in the local Universe. These issues have been the 
focus of major research efforts, largely triggered by the 
Hubble Deep Field observations (Williams et al. 1996) and 
the 10m class telescope data which are becoming available 
now. However, despite the efforts made so far, fundamental 
questions, such as when, where and how galaxies form and 
evolve remain controversial. For instance, in hierarchical 
models for galaxy formation an individual galaxy may pass 
through various phases of disk or spheroid during its 
lifetime (Baugh et al.1996). The authors claim that about 
50\% of all ellipticals (but only about 15\% of spirals) 
would have undergone a major merger during the redshift 
interval 0.0 $\le z \le 0.5$. However, the high degree of 
uniformity in ellipticals star formation history indicates 
that these objects must have formed at z $\ge$ 2 (Bower 
et al. 1992). 

An essential step towards answering these questions is 
understanding the properties of interacting galaxies in 
the local universe and comparing it with the more distant 
objects. We have started a long-term project focusing on 
these issues. Our first step was to select a sample of 
faint interacting galaxies in pairs and groups. By faint 
we mean galaxies that are within $19<m_{R}<22$ and by 
interacting we mean angular separations of $2''<\theta<6''$ 
(see Infante et al. 1996, de Mello et al. 1997a, b for
more details). Among the more interesting results we find 
is the excess of faint pairs at intermediate redshift 
(z$\approx$0.35). 

In this paper we present a continuation of our previous 
work. We present a database of nearby pairs of galaxies 
that will be used in the future to compare with pairs at 
intermediate redshifts.

\section{The Sample}

\subsection{Binary Galaxies}

The ideal sites to study galaxy encounters are pairs and 
groups of galaxies. The pioneers in the studies of pairs of 
galaxies are Holmberg (1937, 1958) and Page (1952, 1961) 
and more recently Karachentsev (1972). Major effort has been 
focused into assembly complete samples of isolated pairs of 
galaxies for the Southern Hemisphere (Reduzzi \& Rampazzo, 
1995, Soares et al. 1995; see Sulentic 1992 for a review on 
the subject). Several studies have analyzed the effect of 
interaction in binary galaxies, e.g. Rampazzo \& Sulentic 
(1992), Rampazzo et al. (1995), de Mello et al.(1996), 
Marquez \& Moles 1996, Reduzzi \& Rampazzo (1996), and de 
Souza et al. (1997).

One simple approach in studying tidal effects is to analyze 
galaxy morphology in binary galaxies. A standard 
morphological classification of binaries is based on their 
components morphology. For instance, early--type binaries, 
usually called EE, are formed by two early--type galaxies 
and late--type binaries (SS) by two spiral galaxies. Mixed 
morphology binaries, ES, are formed by  an early and a 
late--type galaxy. The observed fraction of the three types 
is 0.60 for SS, 0.30 for for ES and 0.10 for EE (Sulentic 
1992). The existence of many isolated mixed binaries raises 
problems for models of galaxy formation that envision local 
environment as a major determinant of morphology. However, 
there are a few evidences, like morphological distortion in 
the late--type galaxy and young stellar population in the 
early--type  showing that mixed binaries are physically 
interacting (de Mello et al. 1996, Rampazzo \& Sulentic 
1992).  Therefore, if mergers of spiral galaxies are the 
mechanism for the formation of ellipticals then binaries 
containing an elliptical would  be the product of 
interaction between at least 3 galaxies. To understand 
whether galaxy transformation is a common phenomenon in 
galaxy evolution  more studies of galaxy morphologies in 
interacting systems are needed. In this work we present the 
results of such analysis for a sample of binary galaxies. 

\subsection{Selection Criteria}

One of the most important goals of this study is to build 
a photometric and spectroscopic database in order to study 
interacting galaxy evolution. For this, the bright {\it and}
faint pairs should constitute a carefully matched sample
with projected separations such that the objects would be 
expected to merge in much less than a Hubble time.  

The faint ($19<m_R<22$) sample was selected from automatic 
compiled catalogs of existing (de Mello et al. 1997a, b) 
CTIO 4m images at the equator; the bright sample 
($16<m_V<18$) was selected from  the catalog of isolated 
binary galaxies (CPG) compiled  by Karachentsev (1972), 
which is the best available catalog that meets the local 
control sample requirements.

Pairs of faint galaxies (separation $\Delta\theta$) were 
chosen such that $\theta_{min} < \Delta\theta < \theta_{max}$. 
Here $\theta_{min}$ is the minimum separation at which 
pairs can be reliably separated (nominally 2$''$); 
$\theta_{max}$ corresponds to some physical separation, $r_p$, 
chosen so that: {\it (i)} physical pairs are doomed to merge 
in $< 10^9$yr (on the basis of both empirical studies and 
conventional dynamics); and {\it (ii)} most pairs in the 
faint sample are physically associated. These conditions are
satisfied by $r_p \simeq 20$ kpc, which corresponds to 
$\sim6''$ at $z = 0.35$.  

What fraction of the faint pairs is expected to be physical 
pairs and merge? Infante et.al. (1996) have calculated the 
angular correlation function, $w(\theta) = 1.75$ for pairs 
at $m_R < 21.5$ and $\theta < 6''$. It follows that 
$\approx 64\%$ of faint pairs will be physically associated.
Furthermore, a cut in $\Delta v$ (say $\Delta v < 1000$ km 
s$^{-1}$) should isolate most of the physical pairs.  
Following Carlberg et.al. (1994) the fraction of galaxies 
that will merge is $70\%$. Thus, it is expected that
$\approx 45\%$ of galaxies in pairs will eventually merge.  

The two samples should be affected by {\it identical} 
selection effects. The CPG contains 603 binaries brighter 
than m$_{pg}$=15.7, is the most complete catalog of binaries 
and has complete redshift information.  In order to match the
properties of z=0.35 pairs we looked at all  Karachentsev 
pairs with characteristics similar to what we expected for 
pairs at high z. We have selected all pairs with projected 
linear separation  $\leq$ 23 h$^{-1}$ kpc 
(h$^{-1}$=100/H$_{0}$kms$^{-1}$Mpc$^{-1}$) and radial 
velocity difference $\leq$ 420 kms$^{-1}$.

\section{Data Reduction and Photometric Calibrations}



B, V and I images were taken at the Cassegrain focus of the 
2.2m MPI/ESO telescope with EFOSC2, CCD19 (1024$\times$1024 
with a scale of 0.336$''$pixel$^{-1}$) during 3 photometric 
nights in la Silla, Chile, in October 1994. Typical exposure 
times were 900s in B, 300s in V and 600s in I. Observational 
conditions were photometric with an average seeing of 1.2$''$. 

Initial reductions followed standard procedures, and were 
performed with IRAF. The bias level was subtracted from 
each image which were then divided by the averaged flat 
field exposure for the night. In figures 
1 and 2 we show the grey-scale V images of the 14 pairs 
in our sample.

\begin{table*}
\begin{flushleft}
\caption[]{Calibration Coefficients}
\small
\begin{tabular}{cccc}
\hline
 & & & \\
\multicolumn{1}{c}{Filter} &
\multicolumn{1}{c}{$m_{0}$}&
\multicolumn{1}{c}{$m_{x}$}&
\multicolumn{1}{c}{$rms$} \\
\hline
     &          &        &        \\
 B   & -0.0027 & 0.2402 & 0.006  \\
 V   & -0.0026 & 0.1531 & 0.012  \\
 I   & -0.0027 & 0.0649 & 0.009  \\
 & & &  \\
\hline
\end{tabular}
\end{flushleft}
\end{table*}

Standard stars from Landolt's list (1992), MarkA, T Phe and 
SAO 98 were observed for extinction and calibration purposes.
To calibrate the images we used the following equations:

\begin{eqnarray}
b - B = A_{1} + A_{2}X + A_{3}(B-V)
\end{eqnarray}
\begin{eqnarray}
v - V = C_{1} + C_{2}X + C_{3}(B-V)
\end{eqnarray}
\begin{eqnarray}
i - I = D_{1} + D_{2}X + D_{3}(V-I)
\end{eqnarray}
\begin{eqnarray}
b - v = E_{1} + E_{2}X + E_{3}(B-V)
\end{eqnarray}
\begin{eqnarray}
v - i = F_{1} + F_{2}X + F_{3}(V-I)
\end{eqnarray}
where B, V and I are the standard magnitudes, b, v and i 
are the instrumental magnitudes, X is the airmass of the 
observation and the coefficients A, C, D, E and F are the 
transformation coefficients. B was obtained from the relation 
$B = b + m_{0} - m_{x}X$, where  m$_{0}$ is 
$m_{0} = -A_{1} - A_{3}((b-v) - E_{1})/E_{3}$ 
and $m_{x} = (A_{2} - A_{3}E_{2}/E_{3})$. Similar relations 
were derived for V and I. 

In Table 1 we present the values of the extinction and color 
coefficients, $m_{x}$ and $m_{0}$, and the rms of the fit 
obtained for each filter. Mean extinction coefficients 
provided by ESO were used for comparison.

Two-dimensional surface photometry was accomplished using 
the last version of GALPHOT image analysis package 
developed and kindly provided by W. Freudling. This package 
runs in the IRAF environment and uses the ellipse--fitting 
routine available in the ISOPHOTE package of STSDAS 
following the method described by Jedrzejewski (1987). 
Prior to ellipse fitting, the sky background is removed 
and field stars and other features are automatically 
masked outside the outer isophotes of the galaxy.


Stars overlapping galaxies were carefully removed by 
extracting elliptical models which were substituted using an 
interpolation across the region. For pairs with overlapping
galaxies such as CPG018, CPG083, CPG091, CPG099, CPG547, 
CPG577, we used the following procedure: first we built 
a model of the pair component {\bf a}, called {\it model a} 
using the ELLIPSE and BMODEL. Then we subtracted 
this model from the original image, creating an {\it image 2}.
From {\it image 2} we built a model for galaxy {\bf b}, 
called {\it model b}. {\it Model b} was subtracted from 
the original image, creating {\it image 3}, which was then 
used to do photometry of galaxy {\bf a} (see figure 
3a for an example of {\it image 3}). With the same procedure 
we have created {\it images 4, 5}  which was used to do 
photometry of galaxy {\bf b} (see figure 
3b for an example of {\it image 5}). In order to test this method 
we have applied it to synthesized images. Good results were 
obtained when the overlapped area was less than 10\% of the 
total area. The goal of this procedure was to recover the 
overlapping external regions of the galaxies. For the cases 
where there were remaining features in the center of 
the subtracted galaxies, we masked these features before 
doing photometry of the companions. Photometry of CPG018 
and CPG083 should be used with caution since they present 
a common envelope.



The radial surface luminosity profiles of the galaxies 
were fitted by an exponential law of the form

\begin{eqnarray}
\Sigma(r) = \Sigma_{o} exp(-r/\alpha)^{\beta}
\end{eqnarray}

where $\Sigma(r)$ is the surface luminosity density 
(in $L_{\odot}$pc$^{-2}$) corresponding to the surface 
brightness, $\mu$ (in mag arcsec$^{-2}$), and $r$ is the 
major axis radius. The fit parameters are then the central 
surface brightness, $\Sigma_0(\mu_0)$, and the exponential 
scale length, $\alpha$. The value of $\beta$ will depend on 
the morphological type of the galaxy, being 1 for spiral 
disks, and $1/4$ for ellipticals or spiral bulges (de 
Vaucouleurs 1948). The following relations were used in 
order to fit the surface brightness versus radius 
and obtain the fit parameters:

\begin{eqnarray}
\mu_{disk} = \mu_{0} + 1.0857 (r/\alpha)
\end{eqnarray}
\begin{eqnarray}
\mu_{bulge} = \mu_{e} + 8.33 ((r/r_{e})^{1/4}-1)
\end{eqnarray}

where $\alpha$ is the disk scale length and $\mu_{0}$ is the 
disk central surface brightness, $r_{e}$ and $\mu_{e}$ are the 
effective radius and the surface brightness at $r_{e}$
radius and surface brightness at $r_e$. We present the radial 
surface brightness distributions and the exponential fits 
in figures 
4 and 5.

The total magnitudes were calculated by extrapolating the 
luminosity profiles to infinity. For several galaxies, bulge 
and disk components were deconvolved from the surface 
brightness profile following a standard interactive procedure 
(Boroson 1981, Schombert \& Bothun 1987, Marquez \& Moles 
1996). First, we have fitted the outer regions of the surface 
brightness profile using an exponential law. We have 
subtracted the fitted disk profile from the total profile 
obtaining the bulge profile. Then, the bulge profile was 
fitted using an $r^{1/4}$ law. Finally, the fitted profiles
(disk and bulge) were summed and the total fitted profile 
was compared with the total observed profile. After some 
iteractions the method converges and we obtain the 
lowest $\chi^2$ fit. The total magnitudes correspond to 
the lowest $\chi^2$. The major uncertainty on the total 
(asymptotic) magnitude values depends on the error on 
the surface brightness ($\mu$) calculated at the outermost 
point reached by the profile determination (due to the 
uncertainty on the sky level) and on the error due to 
the extrapolation procedure. Taking into account all 
contributions, the total uncertainties are 0.18, 0.13 
and 0.12 for B, V and I, respectively. The total magnitudes 
B, V and I,  were corrected for inclination following 
Burstein (1979), Binggelli (1980) and Franx et al. (1991) 
for the ellipticals. For the spirals we have assumed round 
disks. We have also corrected for galactic extinction 
using Burstein \& Heiles (1984) sample. For the galaxies 
that are not in their sample we have followed the method 
in RC2 (de Vaucouleurs et al. 1976). 


In order to check the quality of our photometry we have 
compared our magnitudes with m$_{B}$ in the RC3 (de 
Vaucouleurs et al. 1991) and with other papers in the 
literature, specially with Marquez \& Moles (1994 and 1996, 
MM hereafter) values since our magnitude determination 
followed similar procedure. In figure 
6 we plotted the values of our B magnitudes  before corrections 
(galactic extinction correction and inclination correction)
and m$_{B}$ from RC3. CPG575b was already found as 
discrepant by MM and our result is closer to MM value
than to RC3 (RC3 m$_{B}$= 15.08, MM m$_{B}$= 14.67, 
our value m$_{B}$ = 14.14). CPG580a is discrepant in 
relation to RC3 by 0.86 magnitudes, but that could be due 
to presence of a warped disk which makes the photometric 
decomposition not straightforward. The other two discrepant 
galaxies are CPG587a and b. The distortion on the disks could
also be responsible for this large difference in magnitudes 
(0.83 for CPG587a  and 2.00 for CPG587b), however there is 
no additional magnitude determination in the literature for 
comparison. The other six galaxies that we have in common 
with RC3 have differences in magnitude $\le 0.23$. 
CPG575a, for instance, is only 0.01 fainter in our 
determination than in RC3 and agrees with MM within 
0.11 magnitudes, when no corrections are applied.

A few galaxies in our sample were studied by other authors.
Reshetnikov et al. (1993) has done photometry of CPG099, 
CPG587 and CPG575 however, we can not compare directly the 
total magnitude values because they have used another 
photometric system (the R Cousins system). Rampazzo et al. 
(1995) studied CPG548 but they did not provide total 
magnitude values, only surface brightness profiles. 
They have remarked that the photometric profile of CPG548b 
(NGC~6964) follows an $r^{1/4}$ law, indicating that this 
object is a bona fide elliptical which is in agreement with 
our results for this galaxy. Heraudeau \& Simien (1996) have 
done the photometry of CPG548a galaxy extrapolating the 
brightness distribution by an exponential function. They 
obtained total magnitude values equal to 12.06 and 11.24 for 
V and I bands, respectively. Our total magnitude values 
for this object are different from these values maybe due 
to the differences in the decomposition method and the 
extrapolation procedure. 

\section{Data Analysis}
 
Table 2 lists the data for 14 Karachentsev's pairs observed. 
Column (1) gives the identification in Karachentsev's catalog 
of isolated pairs, column (2) gives other identifications, 
column (3) gives the pair position (1950.0). Columns (4) to 
(6) give the integrated magnitudes in B, V, I calculated as 
described in the previous section.  (7) and (8) give 
the colors B--V and V--I. Column (9) gives the the morphology 
of each galaxy obtained from NED
\footnote{The NASA/IPAC Extragalactic Database, NED, is 
operated by the Jet Propulsion Laboratory, California 
Institute of Technology, under contract with the National 
Aeronautics and Space Administration} ($\rightarrow$ gives 
the new morphology found from the radial profiles). Column 
(10) gives the radial velocities (km s$^{-1}$ obtained from 
NED). 

\begin{table*}
\begin{flushleft}
\caption[]{Data}
\small
\begin{tabular}{rllrrrcclr}
\hline
 & & & & & & & \\
\multicolumn{1}{c}{Name}                  &
\multicolumn{1}{c}{Id.}                   &
\multicolumn{1}{c}{RA$^{h\,m\,\,s}$ DEC$^{\circ}$$'$}&
\multicolumn{1}{c}{B}                     &
\multicolumn{1}{c}{V}                     &
\multicolumn{1}{c}{I}                     &
\multicolumn{1}{c}{B--V}                   &
\multicolumn{1}{c}{V--I}                   &
\multicolumn{1}{c}{Morph.}                &
\multicolumn{1}{c}{$V_{r}$}               \\
\hline
 & & & & & & & & & \\
    CPG018a & UGC 00496  & 00 45 54 +01 05.0 &15.11 &
14.15 &12.71 & 0.96 & 1.44 &E &18648 
    \\
    CPG018b & UGC 00496  & 00 45 54 +01 05.0 &14.78 &
13.80 &12.57 &0.98 &1.23 & E &19003\\
 & & & & & & & & & \\           
    CPG069a & IC0233     & 02 29 03 +02 35.5 &14.24 &
14.15 &13.44 &0.09 &0.71 & dIn & 8266 \\                   
    CPG069b & CGCG388-036& 02 29 03 +02 35.5 &15.82 &
15.59 &14.63 &0.23 &0.96 & Sab & 8303 \\                   
 & & & & & & & & &\\    
    CPG083a & NGC1143    & 02 52 39 --00 23.0 &14.20 &
13.29 &12.03 &0.91 &1.26 & E $\rightarrow$ S0 & 8514 \\        
    CPG083b & NGC1144    & 02 52 39 --00 23.0 &13.23 &
12.14 &11.20 &1.14 &0.94 & Sring & 8714\\                
 & & & & & & & & &\\    
    CPG091a & CGCG391-011& 03 32 45 --01 23.5 &15.26 &
14.55 &13.35 &0.71 &1.20 & Sa & 10674 \\                    
    CPG091b & CGCG391-012& 03 32 45 --01 23.5 &14.82 &
13.78 &12.47 &1.04 &1.31 & Edust& 11086 \\       
 & & & & & & & & &\\    
    CPG099a & NGC1587    & 04 28 06 +00 33.0 &12.29 &
11.50 &10.25 &0.79 &1.25 & E & 3800\\                  
    CPG099b & NGC1588    & 04 28 06 +00 33.0 &13.35 &
12.46 &11.11 &0.89 &1.35 & Epec $\rightarrow$ S0& 3495\\       
 & & & & & & & & &\\    
    CPG545a & UGC11567   & 20 25 42 +00 19.5 &14.08 &
13.12 &11.94 &0.96 &1.18 & SB?& 5762\\                    
    CPG545b & CGCG373-015& 20 25 42 +00 19.5 &13.85 &
12.69 &11.19 &1.16 &1.50 & Sb & 5732\\            
 & & & & & & & & &\\    
    CPG547a & UGC11593   & 20 32 12 +07 48.0 &14.68 &
14.11 &12.90 &0.57 &1.21 & E & 7856 \\               
    CPG547b & UGC11593   & 20 32 12 +07 48.0 &12.67 &
12.48 &11.45 &0.19 &1.03 & Sc & 7887\\                   
 & & & & & & & & &\\    
    CPG548a & NGC6962    & 20 44 51 +00 07.5 &      &
11.07 &10.58 &     &0.48 & SAB(r)ab& 4419\\                   
    CPG548b & NGC6964    & 20 44 51 +00 07.5 &      &
12.20 &10.88 &     &1.32 & E & 4032\\           
 & & & & & & & & &\\   
    CPG551a & UGC11657   & 20 57 12 --02 04.0 &14.00&
13.49 &12.40 &0.51 &1.09 & Pec & 6033 \\                   
    CPG551b & UGC11658   & 20 57 12 --02 04.0 &14.50 &
13.84 &13.02 &0.56 &1.09 & SAB(rs)s& 6025 \\          
 & & & & & & & & &\\   
    CPG575a & NGC7469    & 23 00 46 +08 36.9 &12.86 &
12.65 &12.56 &0.21&0.29 &SAB(rs)s& 5083\\                     
    CPG575b & IC5283     & 23 00 46 +08 36.9 &13.72 &
12.98 &11.71 &0.74 &1.27 & SA(r)cdpec& 5220 \\
                      & & & & & & & & &\\  
    CPG577a & CGCG406-013& 23 08 18 +08 52.0 &14.92 &
14.18 &12.95 &0.74 &1.23 & E & 11646 \\                    
    CPG577b & CGCG406-013& 23 08 18 +08 52.0 &14.94 &
14.01 &12.74 &0.93 &1.27 & E $\rightarrow$ S0?& 11762 \\          
& & & & & & & & &\\  
    CPG580a & NGC7587    & 23 15 24 +09 23.5 &13.35 &
12.06 &10.92 &1.29 &1.14 & SBab& 8819 \\                      
    CPG580b & CGCG406-051& 23 15 24 +09 23.5 &14.53 &
14.03 &12.55 &0.50 &1.48 & S & 8508 \\                      
 & & & & & & & & &\\  
    CPG583a & CGCG380-058& 23 24 00 +02 02.5 &14.46 &
13.54 &12.23 &0.92 &1.31 & S0? $\rightarrow$ S0 & 5876 \\                      
    CPG583b & CGCG380-059& 23 24 00 +02 02.5 &14.46 &
13.54 &12.23 &0.92 &1.31 & Sa  & 5561 \\          
 & & & & & & & & &\\  
    CPG587a & NGC7714    & 23 33 45 +01 52.7 & 11.86 &
11.99 &11.99&-0.13&1.16 & SB(s)b & 2980\\
    CPG587b & NGC7715    & 23 33 45 +01 52.7 & 11.81 &
12.14 &10.22&-0.33&1.92 & Impec& 2933\\

 & & & & & & & & &\\
\hline
\end{tabular}
\end{flushleft}
\end{table*}

\subsection{Morphological Properties}

CPG018 is a very interesting EE pair with a common envelope. 
Elliptical models were fitted and surface brightness 
profiles follow an r$^{1/4}$ law, however the ellipticity 
of the isophotal contours vary from 0 to 0.4 outside 10 
arcsec. 

CPG069 is a late-type pair with the two galaxies showing 
tidal effects in their shapes. CPG069a is a warm IRAS galaxy 
with flux ratio $F_{60}/F_{30} = 2.86$ and its spectra is 
typical of an HII region galaxy (Kailey \& Lebofsky 1988).
CPG069a surface brightness profile in the central regions 
are not well fitted.

CPG083 is a strongly interacting mixed pair also known as 
Arp 118 or VV331. Both galaxies, NGC 1143(E) and NGC1144 (S), 
show starburst activities and signs of tidal distortion. The 
spiral galaxy presents a knotty ring with a peculiar structure 
which can be interpreted as the product of an interaction of 
two disk galaxies (Hippelein 1989). From the surface 
brightness profile it is clear that NGC 1143 is in fact a 
lenticular galaxy.

CPG091 is a mixed pair with moderate signs of interaction. 
The elliptical galaxy presents a strong dust lane and its 
surface brightness profile follows an r$^{1/4}$ law. The 
ellipticity of the isophotal contours is $\approx$ 0.4 outside 
10 arcsec. 

CPG099 is an early-type pair strongly interacting. The two 
early-type galaxies (NGC1587/88) show isophotal distortion 
and asymmetric light distribution. See Borne \& Hoessel (1988) 
for a simulation of this pair. From the surface brightness 
profile it is clear that NGC 1588 is in fact a lenticular 
galaxy. For NGC 1587 the ellipticity of the isophotal contours 
is $\approx$ 0.3 outside 15 arcsec. 

CPG545 is a late-type pair composed by a barred spiral and an 
Sb galaxy. Photometry of this pair is affected by bright 
stars in the field.

CPG547 is a mixed pair in close contact. The barred spiral 
galaxy is edge-on and overlaps the elliptical (Keel 1993). The 
surface brightness profiles of the elliptical galaxy follow 
an r$^{1/4}$ law. The ellipticity of the isophotal contours 
is quite irregular varying from $\approx$ 0 to 0.2 outside 
10 arcsec. 

CPG548 is a mixed pair. The spiral galaxy has two long 
symmetric arms and a weak central bar and the elliptical has 
an elongated shape. The surface brightness profiles follow 
an r$^{1/4}$ law. Rampazzo et al. (1995) studied its 
kinematics and found signs of interaction. The ellipticity of 
the isophotal contours of NGC6964 is $\approx$ 0.3 outside 
20 arcsec.

CPG551 is a late-type pair strongly disturbed by tidal 
effects. 

CPG575 is a late-type pair of galaxy with one galaxy (IC5283) 
very disturbed by tidal effects and the other one (NGC7469) 
with an outer and an inner ring (Buta \& Crocker 1993, Marquez 
\& Moles 1994). Structures can be seen in their surface 
brightness profiles. Seyfert activity is present (Marquez 
\& Moles 1994)

CPG577 is an early-type pair. No common halo was found. The 
surface brightness profiles show that CPG577b is a lenticular 
galaxy. The surface brightness profiles of CPG577a follow an 
r$^{1/4}$ law. The ellipticity of the isophotal contours 
is quite irregular, varying from $\approx$ 0.4 to 0.2 in 
the inner region between 6 to 10 arcsec, and from $\approx$ 
0.2 to 0.4 outside 10 arcsec.

CPG580 is a late-type pair with one of the galaxies very 
disturbed and with a barred spiral with a dust lane and a 
warped disk (NGC7587). 

CPG583 is a mixed pair, with a spiral and a lenticular 
without any clear peculiarity.

CPG587 is a late-type pair with two very disturbed late--type 
galaxies. NGC7715 has an irregular shape and NGC7714 has a bar 
and a peculiar ring.

We have classified these pairs according to the degree of 
distortion that each galaxy presents. We used a slightly 
modified version of the classification scheme proposed by 
Dahari (1985). Pairs with galaxies that showed strong signs 
of interaction were called $\it I$ which corresponds to 
classes 4, 5 and 6 of Dahari. Pairs with moderate signs of 
interaction were called $\it M$ (2 and 3 in Dahari's 
scheme) and pairs with weak or no sign of interaction were 
called $\it W$. Pairs that had one galaxy disturbed
and one normal were classified as M--W. In Table 3 we 
present the classification of the peculiarities 
according to this scheme, the angular separation 
($\theta$($'$)), the projected separation (in h$^{-1}$ kpc)
and the radial velocity difference (in kms$^{-1}$).

We have looked for star formation regions in the galaxies of 
our sample. Color maps (B--V and V--I) were used to identify
these regions. Blue regions, blue tails, bars, patchy and 
knotty features were identified in most of the late--type 
galaxies. The strongest tidal damage and star formation 
regions are found in CPG083 and CPG551. CPG083 is a mixed pair 
with late--type galaxy completely disrupted and with 
prominent blue star formation regions (see also Hippelein 
1989). CPG551a presents several blue knots typical of star 
formation activity and CPG551b presents a blue tail and a 
blue central bar. We have used color maps in order to 
identify dust regions and found clear signs of dust in 
early and late--type galaxies. CPG091b has the strongest 
equatorial dust lane of our sample. The warped galaxy 
CPG580a also shows a strong dust feature. Another interesting 
result is the number of barred galaxies found in our sample. 
Six late--type galaxies are barred, however no late--type 
pair was found to have two barred galaxies. Reduzzi \& 
Rampazzo (1996) also found a frequent presence of bars in 
pairs and suggested that bars would be an efficient transient 
mechanism to transfer gas to the nuclear region and produce
nuclear activity. 

We have also looked for inner substructures in the 
early--type galaxies in our sample following the 
same procedure used in de Mello et al. (1995, 1996)
and Reduzzi et al. (1994). Elliptical models were 
subtracted from the images and residual images 
were created using the ELLIPSE task (IRAF). In this 
procedure the residual intensity along each ellipse
are parametrized in terms of harmonic components.
The deviations from a pure ellipse are expressed
as a Fourier series such that the fourth cosine term 
(C4) describes whether the isophotes are pointy 
(positive) or boxy (negative). After careful residual 
images analysis we verify that six early--type galaxies 
show residual features indicating deviations from 
elliptical isophotes. On the other hand, for the 
remaining five early--type galaxies in our sample 
(CPG018b, CPG083a, CPG099a, CPG547a and CPG548b)
no substructure is observed, and the residual images
of these galaxies are very similar to E506-01 shown 
in Reduzzi et al. (1994).

The images of CPG018a,  CPG091b and CPG577a, 
after subtraction, show ripples and asymmetries 
like AM2312-511 (Reduzzi et al. 1994). The 
model--subtracted image of CPG099b presents a 
substructure, like a cross, elongated towards 
the tidal deformation (south) of the galaxy, similar 
to the E5720420 (Reduzzi et al. 1996). They
classified this residual feature as ``discy isophotes".
According to Reduzzi et al. (1996b), after model
subtraction, discy ellipticals show a  filament 
of light which corresponds to the area where the model
underestimates the light. The filament could be 
real or artificial; the later produced by the 
combined effects of pointy isophotes and twisting 
of the isophotes. The calculated average value of 
the coefficient C4 between $10''$ and $20''$ for 
this galaxy is -0.02. The residual images of 
the CPG577b and CPG583a show the most interesting 
inner features in our sample. CPG577b have 
inner shells like the residual map of NGC~2300 
(Forbes \& Thomson 1992, Plate 2a) and CPG583a 
shows an X-structure similar to the E556-130 (Reduzzi
et al. 1994). Deviations, such as X-structures, 
could be associated to the fact that we were 
adjusting an elliptical model to a boxy galaxy.
Reduzzi et al. (1996b) concluded that in E556-130 
(NGC~2211), the X-structure is described by sudden 
variation in the trend of C4 from a boxiness of 
up to -1.6 percent to a diskiness which reaches 
2 percent. However, Forbes \& Thomson (1992) have
suggested that a variety of phenomena could be 
responsible for the boxy isophotes. We verified
a similar variation of C4 as a function of 
semi-major axis for both CPG577b (-0.007 inner 
$15''$ to 0.02 out) and CPG583a (-0.01 inner $12''$ 
to 0.03 out).

\begin{table*}
\begin{flushleft}
\caption[]{Morphologies}
\small
\begin{tabular}{rlllcrl}
\hline
 & & & & &\\
\multicolumn{1}{c}{Name}                  &
\multicolumn{1}{c}{Type}                  &
\multicolumn{1}{c}{Int.Class  }           &
\multicolumn{1}{c}{$\theta$}              &
\multicolumn{1}{c}{Sep.}                  &
\multicolumn{1}{c}{$\Delta V$}            \\
\multicolumn{1}{c}{} &
\multicolumn{1}{c}{} &
\multicolumn{1}{c}{} &
\multicolumn{1}{c}{($'$)} &
\multicolumn{1}{c}{(h$^{-1}$ kpc)} &
\multicolumn{1}{c}{(kms$^{-1}$)} &
\multicolumn{1}{c}{} \\
 & & & & & &\\
\hline
 & & & & & &\\
     CPG 018 & EE & $\it M$ & 0.34 & 18.6 & 355 & \\
     CPG 069 & SS & $\it I$ & 0.70 & 16.9 & 37 & \\
     CPG 083 & ES & $\it I$ & 0.71 & 17.8 & 200 & \\
     CPG 091 & ES & $\it M-W$ & 0.72 & 22.8 & 412 & \\
     CPG 099 & EE & $\it I$ & 0.97 & 10.3 & 305 & \\
     CPG 545 & SS & $\it W$ & 0.34 & 05.7 & 30 & \\
     CPG 547 & ES & $\it I$ & 0.67 & 15.3 & 31 & \\ 
     CPG 548 & ES & $\it M-W$ & 1.76 & 21.6 & 387 & \\
     CPG 551 & SS & $\it I$ & 0.98 & 17.2 & 8 & \\
     CPG 575 & SS & $\it I$ & 1.32 & 19.8 & 137 & \\
     CPG 577 & EE & $\it W$ & 0.47 & 16.0 & 116 & \\
     CPG 580 & SS & $\it I$ & 0.82 & 20.7 & 311 & \\
     CPG 583 & ES & $\it W$ & 1.00 & 16.6 & 315 & \\
     CPG 587 & SS & $\it I$ & 1.88 & 16.2 & 47 & \\
        
 & & & & & & \\
\hline
\end{tabular}
\end{flushleft}
\end{table*}

\section{Discussion}

The colors of pairs members follow the so-called Holmberg 
effect which is a correlation between the colors of the pairs 
members (Holmberg 1954, Demin et al. 1984). In figure 
7 we show the (B--V) plot for each component (a, b) 
of the pairs. Late-type pairs are marked as filled circles, 
early-type pairs as open circles and mixed-pairs as stars. 
Reduzzi \& Rampazzo (1995) using a more complete sample found 
that while the colors of the members correlate significantly 
for EE and SS pairs, the correlation for ES is less
significant. It suggests that pairs of different morphologies 
should be analyzed differently. In pairs with two gas-rich 
components (SS) tidal forces perturb the gas and the stars 
over the entire disk causing severe damage to both galaxies. 
SS pairs such as CPG551 and CPG587 have strong tails probably 
caused by the slow encounter between the galaxies. Other SS 
pairs have barred galaxies probably also created by tidal 
effects. However, our sample is too small to allow 
any statistical analysis.


Four of our pairs are now confirmed to have a lenticular 
galaxy as a component. Rampazzo et al. (1992 and 1995) 
had also found an excess of lenticulars in their sample. 
In fact, they had estimated that 10-15\% of all pairs 
classified as mixed pairs would in fact be disky pairs, 
i.e. formed by a spiral galaxy and a lenticular galaxy. 
This result suggests that S0's could be the result of 
interaction. Charlton et al. (1992) have proposed a 
toy-model for evolution of morphological types in a 
merger scenario which also agrees with our results. 
They have considered that mergers between unequal mass 
galaxies make S0's and mergers between equal mass 
galaxies make ellipticals. In their model S-S0 pairs 
are formed from an initial group of 3-5 galaxies. 

The distortion on the isophotal shapes of elliptical 
galaxies are frequently used in studies of environmental 
effects as evidences of tidal effects. For instance,
Bender (1988) has suggested that elliptical galaxies 
can be classified according to their flatness and Bender 
et al. (1989) suggested that boxy isophotes are signs of 
past merger (Bender et al. 1989). Caon \& Einasto (1995) 
have suggested that this property is more sensitive to 
the environment rather than the bulge/disk ratio. They 
found that disky elliptical galaxies are located in lower 
density environments and boxy ellipticals in higher 
densities. However, more recently, Andreon (1996) using 
a more complete sample of galaxies in the Coma cluster 
has suggested that diskiness and boxiness has no 
environmental origin. However, because our sample of 
galaxies is made of isolated pairs and that galaxies 
in clusters are more influenced by the cluster as a 
whole than by neighboring galaxies (Makino \& Hut 1997) 
it is possible that tidal effects on the internal 
properties of galaxies be more effective in pairs than 
in clusters. In this study we verified that 6 galaxies
show deviations from their elliptical isophotes or inner 
substructures; the most interesting results were 
obtained for 2 S0 galaxies. One of them, CPG577b, 
shows inner shells and the other one, CPG583a, exhibits 
X-structures. However, because of the small size of our
sample we are unable to draw any conclusion on the 
enviromental origin of these features. However, our 
results agree with previous ones and show that
X-structures is one of the less frequently observed 
forms of fine structure (less than 10\% of our sample 
show this structure).

It is well known that interaction can trigger both 
thermal and non-thermal activity, i.e. starburst and 
AGN (Dahari 1985, Gallimore \& Keel 1993, Keel and van 
Soest 1992, Liu \& Kennicutt 1995). However, while many 
active galaxies show signs of interactions, not all 
interacting galaxies show signs of activity. An intuitive 
solution for this dichotomy is the fact that tidal damage 
is long-lived (few Gyr) while any form of activity is 
short lived (10$^{8}$ yr). Keel et al. (1985) showed 
that Seyfert activity is over-represented in early stage 
interactions. Liu \& Kennicutt (1995) work on merging 
systems found a very large range of spectral properties, 
ranging from completely evolved stellar populations to 
starburst and post-starburst systems. However, samples 
of galaxies such as the one selected by Bergvall \& 
Johansson (1995) lack of such effects (N. Bergvall 
private communication). This is may be due to the fact 
that Bergvall \& Johansson's sample includes weak cases 
of interaction and pairs with large separations, 
which is particularly different from the sample that 
we are analyzing. Some pairs of galaxies of our sample 
are in stages of interaction which present tidal damage 
and also activity. There is strong tidal damage on 
late-type galaxies such as CPG083, CPG551, CPG580 and 
even on early-type galaxies such as CPG099. There is 
Seyfert activity in CPG083 and CPG575 (Hippelein 1989 
and Marquez \& Moles 1994) and several star formation 
regions in CPG551a and CPG580b. However, one has to take 
into account that pairs of different morphologies may 
have passed to different evolution processes and
therefore show different effects. Moreover, activities 
in disturbed disks have a longer lifetime due to a 
larger available gas supply in a disturbed disk (Keel 
1996).

\section{Conclusions}

We have selected a local sample of pairs of galaxies with the 
aim of building a photometric database which will be used in 
the future for comparing local pairs properties with more 
distant pairs. A few properties of local pairs that we have 
discussed in this paper are: 

1) galaxies in close pairs are in general bluer than isolated 
galaxies

2) close pairs of galaxies have an excess of early--type 
galaxies

3) Late--type pairs show strong tidal damage to their 
morphology and blue star formation regions

4) many elliptical galaxies in close pairs are in fact 
lenticular galaxies (2 lenticular galaxies have blue star 
formation regions)

5) pairs with smaller radial velocity difference present 
stronger signs of interaction

6) pairs can present activity and tidal damage simultaneosly

Moreover, one should take into account that local pairs of 
different morphology may have passed to a different evolution 
process which violently transformed their morphology. Pairs 
with at least one early-type component may be descendents of 
groups of galaxies, i.e. the early--type galaxy been already 
a result of a collapse of a group of galaxies. However, 
late--type pairs are probably long--lived showing clearly 
signs of interaction. Some of them could be seen as an early 
stage of mergers because they are still separated but with 
their disks very disturbed. One has also to account for the 
fact that more distant pairs could be in the process of
formation and would have different properties when compared 
to these local pairs. 

\begin{acknowledgements}
 
SJ and DFM thank CNPq for the fellowship and LI thanks 
Fondecyt Chile for support through `Proyeto FONDECyT 8970009' 
and from a 1995 Presidential Chair in Science. We are 
grateful to M.V. Alonso, N. Bergvall, W.C. Keel and J.M. 
Schwartzenberg for helpful comments and W. Freudling for 
providing GALPHOT. DFM thanks the hospitality at Universidad 
Catolica and ESO, Chile, and UNESCO for the international 
grant.

\end{acknowledgements}

\end{document}